\documentclass[11pt]{article}
\usepackage{amsmath,amssymb,color,graphics,epsfig}
\usepackage{amsfonts}
\makeatletter
\@addtoreset{equation}{section}
\makeatother

\newcommand{\be}{\begin{equation}}
\newcommand{\ee}{\end{equation}}
\newcommand{\bea}{\setlength\arraycolsep{2pt} \begin{eqnarray}}
\newcommand{\eea}{\end{eqnarray}}
\newcommand{\nn}{\nonumber}

\def\ft#1#2{{\textstyle{\frac{\scriptstyle #1}{\scriptstyle #2} } }}
\def\fft#1#2{{\frac{#1}{#2}}}

\def\0{{\sst{(0)}}}
\def\1{{\sst{(1)}}}
\def\2{{\sst{(2)}}}
\def\3{{\sst{(3)}}}
\def\4{{\sst{(4)}}}
\def\5{{\sst{(5)}}}
\def\6{{\sst{(6)}}}
\def\7{{\sst{(7)}}}
\def\8{{\sst{(8)}}}
\def\sst#1{{\scriptscriptstyle #1}}

\begin{document}

\begin{flushright}
%\hfill{KIAS-P12028}
 %\hfill{
%\bf hep-th/yymmnnn}
\end{flushright}

\vspace{25pt}
\begin{center}
{\large {\bf Thermodynamical First Laws of Black Holes in \\ Quadratically-Extended Gravities }}

\vspace{10pt}
Zhong-Ying Fan, H. L\"u

\vspace{10pt}

{\it Department of Physics, Beijing Normal University, Beijing 100875, China}

\vspace{40pt}

\underline{ABSTRACT}
\end{center}

Einstein gravities in general dimensions coupled to a cosmological constant and extended with quadratic curvature invariants admit a variety of black holes that may asymptote to Minkowski, anti-de Sitter or Lifshitz spacetimes. We adopt the Wald formalism to derive an explicit formula for calculating the thermodynamical first law for the static black holes with spherical/toric/hyperbolic isometries in these theories.  This allows us to derive/rederive the first laws for a wide range of black holes in literature.  Furthermore, we construct many new exact solutions and obtain their first laws.

\vfill {\footnotesize Emails: zhyingfan@gmail.com \ \ \ mrhonglu@gmail.com}

\thispagestyle{empty}

\pagebreak

\tableofcontents
\addtocontents{toc}{\protect\setcounter{tocdepth}{2}}

%%%%%%%%%%%%%%%%%%%%%%%%%%%%%%%%%%%%%%%%

%\newpage
%%%%%%%%%%%%%%%%%%%%%%%%%%%%%%%%%%%%%%%%

%\vspace{2cm}

\newpage
\section{Introduction}

One important development in General Relativity is fixing the precise relation between the surface gravity $\kappa$ of a black hole to the temperature of its evaporation
\cite{Hawking:1974rv,Hawking:1974sw}
\be
T=\fft{\kappa}{2\pi}\,.
\ee
This, together with the earlier work \cite{Bekenstein:1973ur}, enables one to obtain the entropy area law, namely the entropy of a black hole is equal to one quarter of the area of the horizon:
\be
S=\ft14 \hbox{Area}_H\,.
\ee
Whilst the temperature formula turns out to be true in general owing to its robust geometric origin \cite{Gibbons:1976ue}, the entropy formula acquires modifications when gravity is extended by higher-order curvature terms.  In \cite{wald1,wald2}, Wald develops a formalism that allows one to calculate the entropy from a rather general formula, namely
\be
S=-\ft18\int_{+} \sqrt{h} d\Sigma\, \epsilon^{ab} \epsilon^{cd} \fft{\partial L}{\partial R^{abcd}}\,,\label{waldentropy}
\ee
where the integration is over the bifurcation surface of the horizon and $L$ is related to the Lagrangian by ${\cal L}= \sqrt{-g}\, L$. (Here $h$ is the determinant of the metric on the horizon and $\epsilon^{ab}$ is the binormal to the bifurcation surface.)  In many cases, especially for black holes that are asymptotic to the flat spacetime, knowing the temperature, entropy, together with some clever way of deriving the mass and other conserved quantities, is enough for one to obtain the thermodynamical first law of the black holes.

The situation becomes more complicated when one considers black holes that asymptote to anti-de Sitter (AdS) or Lifshitz spacetimes. In higher-order gravities, in addition to the massless graviton modes, there can also be massive scalar or spin-2 modes emerging. With suitable coupling constants, all these modes can arise in AdS or Lifshitz black holes.  Since these additional hairs are not associated with some conserved quantities, there is no obvious way to define them.  They arise simply as integration constants in a solution and there is no reason that they should always be fixed in any thermodynamical setup.  The purpose of this paper is to address how these hairs contribute and modify the thermodynamical first law of the black holes.

The phenomenon already emerges in Einstein-dilaton gravity that admits AdS vacua. Scalar charges in AdS backgrounds were well studied in literature; see, e.g.
\cite{Henneaux1,Henneaux2,Henneaux3,Hertog1,Hertog2,Hertog3}.
However, how these scalar charges modify the thermodynamical first law wasn't addressed until \cite{lupapo}, where the Kaluza-Klein dyonic AdS black hole was constructed in maximal gauged supergravity in four dimensions.  It was observed that the first law of the black hole thermodynamics would be violated unless one included also the scalar charge contribution \cite{lupapo}.\footnote{Alternatively, this phenomenon was interpreted in \cite{Chow:2013gba} as the black hole having no well-defined mass, where a more general class of dyonic AdS black holes were obtained in maximal gauged supergravity.}

Although the black hole temperature and entropy do not have smooth classical limit, the corresponding thermodynamical first law does, since the $\hbar$ in $T$ and $S$ cancels out precisely in $TdS$.  The general Wald formulism, a pure classical calculation, in principle allows one to derive the first law; however, there is no simple formula like (\ref{waldentropy}). Indeed, evaluating the Wald formula on a black hole horizon always gives $TdS$ \cite{wald1,wald2},\footnote{Suitable gauge choice is necessary when it is applicable.} its evaluation on the asymptotic regions depends on the detail of the theory and the structure of the solutions.  For static AdS black holes with spherical/toric/hyperbolic isometries, the first law involving scalar hair \cite{liulu,Lu:2014maa} and massive vector (Proca field) hair \cite{Liu:2014tra} were obtained. The formalism was used to address properly the thermodynamics of Lifshitz black holes \cite{Liu:2014dva}.  The first law of Yang-Mills black holes was also obtained in \cite{Fan:2014ixa}.

In this paper, we consider Einstein gravities with a cosmological constant in general dimensions, extended with three quadratic curvature invariants. The theories admit a variety of black holes that are asymptotic to Minkowski, AdS and Lifshitz spacetimes. We focus on the static solutions with spherical/toric/hyperbolic isometries and study their first laws of thermodynamics.  The paper is organized as follows.  In section 2, we present the general action of the theories and obtain the equations of motion.  In section 3, we obtain the explicit Wald formula for the static solutions involving two metric functions.  In sections 4, 5, we derive the first laws for AdS and Lifshitz black holes respectively.  We test the results with a wide range of exact solutions in literature.  In section 6, we construct new exact solutions and obtain their first laws.  We conclude the paper in section 7.

\section{Gravity extended with quadratic curvature invariants}

The action for Einstein gravity with a cosmological constant extended with quadratic curvature invariants in general $n$ dimensions is given by
\bea
S&=& \int d^nx \sqrt{-g}\, L\,,\cr
L&=&\kappa(R-2\Lambda_0)+\alpha R^2+\beta R_{\mu\nu}R^{\mu\nu}+\gamma R_{\mu\nu\lambda\rho}R^{\mu\nu\lambda\rho}\,,
\label{genaction}
\eea
where $(\kappa,\alpha,\beta,\gamma)$ are coupling constant and $\Lambda_0$ is the bared cosmological constant.  (There should be no confusion of this $\kappa$ with the surface gravity mentioned in the introduction.)  The variation of the action with respect to the metric takes the following form
\be \delta S=\int \mathrm{d}^{n}x\sqrt{-g}[E_{\mu\nu} \delta g^{\mu\nu}+\nabla_\mu J^{\mu}]\,.\label{varyaction}
\ee
The tensorial quantities $E_{\mu\nu}$ and $J^\mu$ are given by
\bea
E_{\mu\nu}&=&\kappa[R_{\mu\nu}-\ft 12 g_{\mu\nu}(R-2\Lambda_0)]+\alpha [2R (R_{\mu\nu}-\ft 14 g_{\mu\nu}R)+2(g_{\mu\nu}\square-\nabla_\mu \nabla_\nu)R]\cr
&&+\beta [(g_{\mu\nu}\square-\nabla_\mu \nabla_\nu)R+\square(R_{\mu\nu}-\ft 12 g_{\mu\nu}R)+2(R_{\mu\lambda\nu\rho}-\ft 14 g_{\mu\nu}R_{\lambda\rho})R^{\lambda\rho}]\cr
&&+\gamma[-\ft 12 g_{\mu\nu}R_{\alpha\beta\gamma\eta}R^{\alpha\beta\gamma\eta} +2R_{\mu\lambda\rho\sigma}R_\nu^{\ \lambda\rho\sigma}+4R_{\mu\lambda\nu\rho}R^{\lambda\rho}\cr
&&\qquad\qquad\qquad -4R_{\mu\sigma}R^\sigma_\nu+4\square R_{\mu\nu}-2\nabla_\mu \nabla_\nu R]\,,\cr
J^{\mu}&=&[(\kappa+2\alpha R)G^{\mu\nu\rho\lambda}+\beta T^{\mu\nu\rho\lambda}+4\gamma R^{\mu\rho\lambda\nu}] \nabla_\nu \delta g_{\rho\lambda} \cr
&&-[2\alpha G^{\mu\nu\rho\lambda}\nabla_\nu R+\beta \nabla_\nu T^{\nu\mu\rho\lambda}+4\gamma \nabla_\nu R^{\mu\rho\lambda\nu}] \delta g_{\rho\lambda}\,,\label{emunujmu}
\eea
where
\bea G^{\mu\nu\rho\sigma} &=&\ft 12 (g^{\mu\rho}g^{\nu\sigma}+g^{\mu\sigma}g^{\nu\rho}) - g^{\mu\nu}g^{\rho\sigma},\cr
T^{\mu\nu\rho\lambda} &=& g^{\mu\rho}R^{\nu\lambda} +g^{\mu\lambda}R^{\nu\rho}-g^{\mu\nu}R^{\rho\lambda}-
g^{\rho\lambda}R^{\mu\nu}\,.\label{gt}
\eea
It is useful to know that the $G$ and $T$ tensors satisfy the following identities
\bea
&&G^{\mu\nu\rho\sigma}=G^{(\mu\nu)(\rho\sigma)},\quad T^{\mu\nu\rho\lambda}=T^{\mu\nu(\rho\lambda)}\,,\cr
&&\nabla_\nu T^{\nu\mu\rho\lambda}=2\nabla^{(\rho}R^{\lambda)\mu}- \nabla^{\mu}R^{\rho\lambda}-\ft12g^{\rho\lambda}\nabla^\mu R,\cr
&&\nabla_\nu R^{\mu\rho\lambda\nu}=\nabla^\rho R^{\mu\lambda}-\nabla^\mu R^{\rho\lambda}\,.
\eea
Since the total derivative term in the variation of the action (\ref{varyaction}) vanishes for appropriate boundary condition, the equations of motion are then
\be
E_{\mu\nu}=0\,.\label{eom0}
\ee

In this paper, we shall focus on the static black holes with the spherical/toric/hyperbolic isometries.  The most general metric Ansatz takes the form
\be
ds^2 = -h(r) dt^2 + \fft{dr^2}{f(r)} + r^2 d\Omega_{n-2,k}^2\,,\label{metricansatz}
\ee
where $k=1,0,-1$ for which $d\Omega_{n-2,k}^2$ is the metric for an $(n-2)$-dimensional round sphere, torus or hyperboloid.  Since the Ansatz is the most general under the symmetries, one can substitute the Ansatz into the action directly and then vary the resulting action with respect to $h$ and $f$ and obtain two equations of motion.  The equations of motion are rather complicated and they do not appear to provide much insight, and hence we shall present only the following quantities that are useful to compute the effective action:
\bea
&&R =\Big( -\fft{h''}{h} + \fft{h'^2}{2h^2} + \fft{h'f'}{2hf} -\fft{n-2}{r}\big(\fft{h'}{h} +\fft{f'}{f}\big) -\fft{(n-2)(n-3)}{r^2}\big(1-\fft{k}{f}\big)\Big)f\,,\cr
&&R_{\mu\nu}R^{\mu\nu}=\ft18\Big(\fft{h''}{h} - \fft{h'^2}{2h^2} + \fft{h'f'}{2hf}\Big)^2 f^2 + \fft{(n-1)(n-2) f^2}{4r^2} \Big(\fft{h'^2}{h^2} +
\fft{f'^2}{f^2}\Big)\cr
&&\qquad\qquad+ \fft{(n-2)f^2}{r}\Big(\fft{h''h'}{2h^2} + \fft{h''f'}{2hf} - \fft{h'^3}{4h^3} + \fft{h'f'^2}{4h f^2} + \fft{h'f'}{2rhf}\Big)\cr
&&\qquad\qquad + \fft{(n-2)(n-3) f^2}{r^3} \Big(1 - \fft{k}{f}\Big)\Big(
\fft{h'}{h} + \fft{f'}{f} + \fft{n-3}{r} \big(1 - \fft{k}{f}\big)\Big)\,,\cr
&&R_{\mu\nu\rho\sigma}R^{\mu\nu\rho\sigma} = \Big(\fft{h''}{h} - \fft{h'^2}{2h^2} + \fft{h'f'}{2hf}\Big)^2 f^2 +
\fft{(n-2)f^2}{r^2}\Big(\fft{h'^2}{h^2} + \fft{f'^2}{f^2}\Big)\cr
&&\qquad\qquad\qquad+ \fft{2(n-2)(n-3)f^2}{r^4} \Big(1 - \fft{k}{f}\Big)^2\,.
\eea
In this paper, a prime denotes a derivative with respect to $r$.  We leave to the interested readers to obtain the two explicit equations of motion for the metric functions $h$ and $f$ themselves.

\section{Wald formalism}

Extended gravities considered in section 2 turn out to provide a variety of vacuum solutions such as Minkowski, (A)dS and Lifshitz spacetimes.  There are also black holes that are asymptotic to these vacua. A very important property of a black hole is its thermodynamics culminated in the form of its first law.  Although the thermodynamics of a black hole is an intrinsic quantum phenomenon, there is a smooth classical limit of its first law.  While the Wald formulism provides a way of calculating the first law for general black holes, it is a rather complicated procedure.  In this section, we shall specialize the Wald formalism to metrics of the type (\ref{metricansatz}) for extended gravities and obtain a formula that one can readily use.

\subsection{Review of the general Wald formalism}

The Wald formalism \cite{wald1,wald2} starts with the variation of the action (\ref{varyaction}).  For the given current $J^\mu$,  one can define a current 1-form and its Hodge dual
\be
J_\1=J_\mu dx^\mu\,,\qquad
\Theta_{\sst{(n-1)}}={*J_\1}\,.
\ee
When the variation is generated by a Killing vector, one can define the Noether current $(n-1)$-form as
\be
J_{\sst{(n-1)}}=\Theta_{\sst{(n-1)}}-i_\xi\cdot {*L}\,,
\ee
where $i_\xi\cdot$ denotes the contraction of $\xi$ to the first index of the tensor it acted upon. It was shown in \cite{wald1,wald2} that once the equations of motion are satisfied, i.e.~$E_{\mu\nu}=0$,  one has
\be
dJ_{\sst{(n-1)}}= 0.
\ee
Thus one can define a charge $(n-2)$-form as
\be J_{\sst{(n-1)}}= dQ_{\sst{(n-2)}}\,.
\ee
It was shown in \cite{wald1,wald2} that the variation of the Hamiltonian with respect to the integration constants of a given solution is:
\be
\delta H=\frac{1}{16\pi}[\delta \int_{\mathcal{C}}J_{\sst{(n-1})}- \int_{\mathcal{C}}d(i_\xi\cdot \Theta_{\sst{(n-2)}})]=\frac{1}{16\pi}\int_{\Sigma_{n-2}}[\delta Q_{\sst{(n-2)}}-i_\xi\cdot \Theta_{\sst{(n-2)}}]\,.\label{generalwald}
\ee
where $\mathcal{C}$ is a Cauchy surface, $\Sigma_{n-2}$ is its two boundaries, one on the horizon and the other at infinity.  For a static or stationary black hole, one lets $\xi$ be the time-like Killing vector that becomes null on the horizon, and the first law of black hole thermodynamics emerges by the vanishing of $\delta H$.  Evaluating $\delta H$ at both asymptotic infinity and on the event horizon yields
\be
\delta H_\infty = \delta H_+\,.
\ee
The Wald formula (\ref{generalwald}) is rather general.  For our quadratically-extended gravities, various quantities in (\ref{generalwald}) are given by
\bea
J_{\sst{(n-1)}} &=&2\varepsilon_{\mu c_1...c_{n-1}} \nabla_\lambda\Big((\kappa+2\alpha R)\nabla^{[\lambda}\xi^{\mu]}+ (4\alpha+\beta)\xi^{[\lambda}\nabla^{\mu]}R\cr
&&+(2\beta+8\gamma)\nabla^{[\mu}R^{\lambda]\sigma}\xi_\sigma+2\beta R^{\sigma[\lambda}\nabla_\sigma \xi^{\mu]}+2\gamma R^{\lambda\mu\sigma\rho}\nabla_\sigma \xi_\rho\Big)\,,\cr
Q_{\sst{(n-2)}} &=& -\varepsilon_{\mu\nu c_1...c_{n-2}}\Big((\kappa+2\alpha R)\nabla^{\mu}\xi^{\nu}+(4\alpha+\beta) \xi^{\mu}\nabla^{\nu}R\cr
&&+(2\beta+8\gamma)\nabla^{\nu}R^{\mu\sigma}\xi_\sigma+2\beta R^{\sigma\mu}\nabla_\sigma \xi^{\nu}+2\gamma R^{\mu\nu\sigma\rho}\nabla_\sigma \xi_\rho\Big)\,,\cr
i_{\xi}\cdot \Theta_{(n-1)}&=&-\varepsilon_{\lambda\mu c_1...c_{n-2}}\xi^{\lambda}J^{\mu}\cr
&=&-\varepsilon_{\lambda\mu c_1...c_{n-2}}\xi^{\lambda}\Big(((\kappa+2\alpha R)G^{\mu\nu\rho\lambda}+\beta T^{\mu\nu\rho\lambda}+4\gamma R^{\mu\rho\lambda\nu}) \nabla_\nu \delta g_{\rho\lambda}\cr
&&-(2\alpha G^{\mu\nu\rho\lambda}\nabla_\nu R+\beta \nabla_\nu T^{\nu\mu\rho\lambda}+4\gamma \nabla_\nu R^{\mu\rho\lambda\nu}) \delta g_{\rho\lambda}\Big)\,.\label{various}
\eea
These specialized formulae however are still rather complicated and they are not in the forms that one can easily substitute a solution into.  In the next subsection, we further specialize to the static solutions with spherical/toric/hyperbolic isometries.

\subsection{Applying for the static solutions}

We now evaluate the general Wald formula (\ref{generalwald}) with (\ref{various}) for the static Ansatz (\ref{metricansatz}).  Let $\xi=\partial/\partial t$, we obtain,  by some tedious calculations, that
\be
\delta H=\frac{\omega}{16\pi}\ r^{n-2}\sqrt{\frac{h}{f}}\,\Big(F_0\delta f+F_1\delta (f')+F_2\delta (f'')+R_0 \delta h+R_1\delta (h')+R_2\delta (h'')+R_3\delta (h''')
\Big)\,, \label{wald1}
\ee
where $\omega$ is the volume factor of the $(n-2)$-dimensional space of $d\Omega_{(n-2),k}^2$ and the functions $F_{0,1,2},\ R_{0,1,2,3}$ are
\bea F_0&=&\frac{n-2}{r}\Big(-\kappa+2(n-3)((n-8)\alpha-\beta)\frac{f}{r^2} +(8(n-3)\alpha+(2n-5)\beta+4\gamma)\frac{f'}{2r}\cr
&&+\ft12(4\alpha+\beta)f''\Big) +(2\alpha+\beta+2\gamma)\frac{fh'}{h}\Big(\frac{f''}{2f}+
\frac{3f'h''}{2fh'}+\frac{7h'^2}{2h^2}+
\frac{3h'''}{h'}-\frac{7h''}{h}-\frac{f'h'}{fh}\Big)\cr
&&+\frac{(n-2)fh'}{2rh}\Big(2(3\alpha+\beta+\gamma)(\frac{f'}{f}-
\frac{3h'}{h})+(16\alpha+7\beta+12\gamma)\frac{h''}{h'}-
\frac{8\alpha+5\beta+12\gamma}{r}\Big)\cr
&&-\fft{2k (n-2)(n-3)(n-4)\alpha}{r^3},\cr
F_1&=&(n-4)(n-2)(4\alpha+\beta)\frac{f}{r^2}+(2\alpha+\beta+2\gamma)
 \Big((n-2)\frac{h'}{rh}-\frac{5h'^2}{2h^2}+\frac{3h''}{h}\Big)f\,,\cr
F_2&=&(n-2)(4\alpha+\beta)\frac{f}{r}+(2\alpha+\beta+2\gamma)\frac{f h'}{h}\,,\cr R_0&=&(n-2)\frac{f^2h'}{2rh^2}\big[(20\alpha+7\beta+8\gamma)\frac{h'}{h}-
(8\alpha+3\beta+4\gamma)\frac{f'}{f}\big]+(2\alpha+\beta+2\gamma)
\frac{f^2h'^2}{h^3}\cr
&&\times \Big(2(n-2)\frac{h(h'-rh'')}{r^2h'^2}-\frac{2hh'''}{h'^2}+ \frac{9f'}{2f}+\frac{9h''}{h'}-\frac{7h'}{h}-\frac{hf''}{h'f}
-\frac{3hh''f'}{h'^2f}\Big)\,,\cr
R_1&=&(n-2)\frac{f^2}{2rh}\Big(-(20\alpha+7\beta+8\gamma)\frac{h'}{h}+
(8\alpha+3\beta+4\gamma)\frac{f'}{f}\Big)+(2\alpha+\beta+2\gamma)
\frac{f^2}{h}\cr
&&\times \Big(\frac{f''}{f}+\frac{7h'^2}{h^2}-\frac{4h''}{h}-
\frac{9f'h'}{2fh}-\frac{2(n-2)}{r^2}\Big)\,,\cr
R_2&=&(2\alpha+\beta+2\gamma)\frac{f^2}{h}\Big(\frac{3f'}{f}-
\frac{5h'}{h}+\frac{2(n-2)}{r}\Big)\,,\qquad
R_3=2(2\alpha+\beta+2\gamma)\frac{f^2}{h}\,.
\eea
Note that the topology-dependent term appears in (\ref{wald1}) only through $F_0$ with the $\alpha$ coefficient.

It is of interest to consider some special combinations of the quadratic curvature terms.  One is the Gauss-Bonnet combination, corresponding to setting
\be
\beta=-4\alpha\,,\qquad \gamma=\alpha\,.
\ee
The formula (\ref{wald1}) is reduced significantly, and it becomes
\be
\delta H= \fft{\omega}{16\pi} r^{n-2}\, \sqrt{\fft{h}{f}}\, \Big(
-\fft{(n-2)\kappa}{r} + \fft{2(n-2)(n-3)(n-4)\alpha (f-k)}{r^3} \Big)\delta f\,.
\ee
The other combination is the Weyl-squared term, corresponding to
\be
\alpha = \fft{2\gamma}{(n-1)(n-2)}\,,\qquad \beta=-\fft{4\gamma}{n-2}\,.\label{ew}
\ee
There is no dramatic simplification for (\ref{wald1}) in this combination.  In four dimensions, when the theory involves only the Weyl-squared term, it becomes conformal gravity.  Up to an overall scaling, the most general Ansatz is now reduced to $h=f$.  The resulting $\delta H$ was obtained in \cite{Fan:2014ixa}.

Although our main result (\ref{wald1}) is not particularly simple per se, one can nevertheless readily substitute any static solution of the form (\ref{metricansatz}) into (\ref{wald1}) and obtain the first law.  We shall demonstrate this in sections 4, 5 and 6.

\subsection{Evaluating the Wald formula on the event horizon}

It was shown in \cite{wald1,wald2} that $\delta H$ evaluated on the event horizon yields a simple generic formula
\be
\delta H_+ = T\delta S\,,\label{waldhor}
\ee
where the temperature $T$ and entropy $S$ are given by
\be
T=\fft{\kappa}{2\pi}\,,\qquad S=-\fft{1}{8} \int_+ \sqrt{h}\, d^{n-2}x\, \epsilon^{ab}\epsilon^{cd}\, \fft{\partial L}{\partial R^{abcd}}\,.
\ee
Again, $\kappa$ here is the surface gravity on the horizon and it should not to be  confused with the coupling constant in the Lagrangian.  For our metric Ansatz, they are given by
\bea T &=&\fft1{4\pi}\sqrt{h'(r_0)f'(r_0)}\,,\cr
S &=& \ft14{\cal A} \Big[\kappa - (n-2)(4\alpha + \beta) \fft{f'(r_0)}{r_0}+
\fft{2k (n-2)(n-3)\alpha}{r_0^2}\cr
&&\qquad -
\ft14 (2\alpha + \beta + 2 \gamma) \Big(f''(r_0) + \fft{3 f'(r_0) h''(r_0)}{h'(r_0)}\Big)\Big]\,,\label{generalts}
\eea
where $r=r_0$ is the location of the event horizon, and ${\cal A}=\omega r_0^{n-2}$ is its area.  In the extremal limit, the temperature vanishes and the entropy becomes
\be
S_{\rm ext} = \ft14 {\cal A} \Big(\kappa - (2\alpha + \beta + 2\gamma) f''(r_0) + \fft{2k (n-2)(n-3)\alpha}{r_0^2} \Big)\,.
\ee
To verify that (\ref{waldhor}) is indeed true for our specialize cases, we perform Taylor expansions near the horizon, namely
\bea
f(r)&=&\tilde{a}_1(r-r_0)+\tilde{a_2}(r-r_0)^2+ \tilde{a}_3(r-r_0)^3+\tilde{a}_4(r-r_0)^4+\cdots\,,\cr
h(r)&=&\tilde{b}_1(r-r_0)+\tilde{b_2}(r-r_0)^2+ \tilde{b}_3(r-r_0)^3+\tilde{b}_4(r-r_0)^4+ \cdots\,.\label{horizontaylor}
\eea
Substituting these expansions into the equations of motion, we find that there are six independent parameters at the horizon, namely $(r_0,\ \tilde{a}_1,\ \tilde{a}_2,\ \tilde{b}_1,\ \tilde{b}_2,\ \tilde{b}_3)$. The rest coefficients can be solved in terms of these six parameters.

Plugging (\ref{horizontaylor}) into (\ref{wald1}), and then taking the $r\rightarrow r_0$ limit, we obtain
\bea
 \delta H_+ &=& \frac{(n-2)\omega r_0^{n-4} \delta r_0}{32\pi}\,\sqrt{\frac{\tilde{a}_1}{\tilde{b}_1}}\,
\Big(2\kappa \tilde{b}_1r_0-2(n-3)(4\alpha+\beta) \tilde{a}_1\tilde{b}_1 \cr
&&\qquad-r_0(2\alpha+\beta+2\gamma) (3\tilde{a}_1\tilde{b}_2+\tilde{a}_2\tilde{b}_1) + 4k (n-3)(n-4)\alpha \tilde b_1\Big)\,.
\eea
On the other hand, it follows from (\ref{generalts}) that we have
\bea
T =\frac{\sqrt{\tilde{a}_1\tilde{b}_1}}{4\pi}\,,&&
S = \frac{\omega r_0^{n-3}}{8\tilde{b}_1}\Big(2\kappa \tilde{b}_1r_0
-2(n-2)(4\alpha+\beta)\tilde{a}_1\tilde{b}_1\cr
&&\qquad-r_0(2\alpha+\beta+2\gamma) (3\tilde{a}_1\tilde{b}_2+\tilde{a}_2\tilde{b}_1)
+ 4k (n-3)(n-4)\alpha \tilde b_1\Big)\,.
\eea
It is now straightforward to verify that (\ref{waldhor}) is indeed valid.

According to the Wald formalism, the thermodynamical first law of a black hole is simply
\be
\delta H_{\infty} = T \delta S\,.\label{wald2}
\ee
There is however no simple general formula for the evaluation of $\delta H$ at the asymptotic infinity and we shall study it in the case-by-case basis in the following sections.

\section{AdS black holes}

\subsection{AdS vacua and asymptotic linearized modes}

The general theory of extended gravity (\ref{genaction}) admits (A)dS vacua with the cosmological constant $\Lambda$, satisfying
\be
2(n-4)\Big(n(n-1)\alpha + (n-1)\beta + 2 \gamma\Big) \Lambda^2 +
\kappa (n-1)(n-2)^2 (\Lambda - \Lambda_0)=0\,.
\ee
The corresponding (A)dS vacuum solutions are given by $h=f= g^2 r^2 + k$, where the parameter $g$ is given by $\Lambda=-\ft12 (n-1)(n-2) g^2$.  In this section, we are interested in black hole solutions that asymptote to AdS vacua with $g^2> 0$.  At the asymptotic $r\rightarrow \infty$ region, the deviation from the AdS vacua can be viewed as the linear excitations of the backgrounds.  For a generic quadratically-extended gravity, there are six independent linear perturbations for the static Ansatz (\ref{metricansatz}), forming three pairs of falloffs.  They correspond to the usual graviton modes, the massive scalar modes and spin-2 modes.  Writing the metric functions as
\be
h=g^2 r^2 + k + h_1\,,\qquad f=g^2 r^2 + k + f_2\,,
\ee
we find that at the linear level, the large-$r$ expansions of $(h_1,f_1)$ are given by
\bea
h_1 &=& \lambda_0 (g^2 r^2 +k) - \fft{m}{r^{n-3}} + \fft{\xi_1}{r^{\fft12(n-5 -\sigma_1)}} + \fft{\xi_2}{r^{\fft12(n-5+\sigma_1)}} + \cdots\,,\cr
f_1 &=& -\fft{m}{r^{n-3}} + \fft{\eta_1}{r^{\fft12(n-5 -\sigma_2)}} + \fft{\eta_2}{r^{\fft12(n-5+\sigma_2)}}\cr
 &&\qquad\qquad+ \fft{(n-1-\sigma_1) \xi_1}{2(n-1) r^{\fft12(n-5 -\sigma_1)}} + \fft{(n-1+\sigma_1)\xi_2}{2(n-1)r^{\fft12(n-5+\sigma_1)}} + \cdots\,,
\eea
where
\bea
\sigma_1^2 &=& \fft{8n(n-1)\alpha + (n-1)(n+7)\beta + 4 (n^2-6n+17)\gamma-4\kappa g^{-2}}{\beta + 4\gamma}\,,\\
\sigma_2^2 &=& \fft{4(n-2)g^{-2}\kappa -(n-1)[4(n^2-6n-1)\alpha -
(n^2-9n+32)\beta] + 4(n^2-6n+17)\gamma}{4(n-1)\alpha + n\beta + 4 \gamma}\,.\nn
\eea
(These are the exact linear solutions when $k=0$.)
It is clear that the parameters $(\lambda_0,m)$, $(\eta_1,\eta_2)$ and $(\xi_1,\xi_2)$ are associated with the graviton, massive scalar and massive spin-2 modes respectively. It is worth noting that the scaling dimensions of the products $\lambda_0 m$, $\eta_1\eta_2$ and $\xi_1 \xi_2$ are all the same.  The parameter $\lambda_0$ represents the freedom of arbitrary constant scaling the time coordinate.  To fix the definition of time at the asymptotic infinity, we shall always set $\lambda_0=0$.

In special cases of the coupling constants $(\kappa, \alpha, \beta,\gamma)$, the massive modes may disappear.  For example, when the quadratic curvature terms form the Gauss-Bonnet combination, both massive scalar and spin-2 modes decouples from the spectrum, giving
\be
h=f= g^2 r^2 + k - \fft{m}{r^{n-3}} + \cdots
\ee
in the large-$r$ expansion.  For the Einstein-Weyl combination, the massive scalar modes decouple, giving
\bea
\tilde h &=& - \fft{m}{r^{n-3}} + \fft{\xi_1}{r^{\fft12(n-5 -\sigma_1)}} + \fft{\xi_2}{r^{\fft12(n-5+\sigma_1)}}\,,\cr
\tilde f &=& -\fft{m}{r^{n-3}} + \fft{(n-1-\sigma_1) \xi_1}{2(n-1) r^{\fft12(n-5 -\sigma_1)}} + \fft{(n-1+\sigma_1)\xi_2}{2(n-1)r^{\fft12(n-5+\sigma_1)}}\,,\cr
\sigma_1^2 &=& (n-3)^2 - \fft{(n-2)\kappa}{(n-3)g^2\gamma}\,.
\eea

From these linearized modes in the large-$r$ expansion, we see that the terms associated with $\xi_1$ and $\eta_1$ always fall slower than the mass term $m$; they can even be divergent, depending on the specific values of the coupling constants.
If a mode has divergence that is in higher order than $r^2$,  it must be excluded from the solution in order to preserve the asymptotic AdS structure.  Then the existence of such a black hole solution requires a delicate fine tuning even if the no-hair theorem can be bypassed. On the other hand, the $\xi_2$ and $\eta_2$ terms can never be more divergent than $r^2$ and hence they will preserve the asymptotic AdS structure; in fact, they can fall faster than the mass term if $\sigma_1$ and $\sigma_2$ are bigger than $(n-1)$.  In general an AdS black hole in extended gravity could involve all these hairs asymptotically.

\subsection{Conjectured first law from the linear analysis}

Having obtain the general structure of a black hole solution asymptotically at the linear level, we can substitute it into our formula (\ref{wald1}). Assuming that there is no term more divergent than $r^2$ in $\tilde h$ and $\tilde f$, we find that $\delta H_\infty$ is convergent, and given by
\be
\delta H_\infty = c_0\,\delta m + c_1\, \xi_1 \delta \xi_2 + c_2\, \xi_2 \delta \xi_1 + c_3\, \eta_1 \delta \eta_2 + c_4\, \eta_2\delta \eta_1\,,\label{adswald1}
\ee
where $(c_1,c_2,c_3,c_4)$ are complicated expressions of $(\kappa,\alpha,\beta,\gamma,n)$ that we do not find instructive to present.  The expression for $c_0$ is rather simple, however, given by
\be
c_0 = \fft{(n-2)\omega}{16\pi} \Big[ \kappa - 2g^2\Big( n(n-1)\alpha + (n-1)\beta - 2(n-4)\gamma\Big)\Big]\,.
\ee
We can now make a Legendre-like transformation, and define
\be
\widetilde M = c_0\, m + c_2\, \xi_1 \xi_2 + c_4\,\eta_1 \eta_2\,,\label{legendre}
\ee
and the expression (\ref{adswald1}) becomes rather simple, giving
\bea
\delta H_\infty &=&\delta \widetilde M + \fft{(n-2)\omega\sigma_1}{(n-1)32\pi}
\big(\kappa g^{-2} - 2(n-1)(\beta + n\alpha)+ 4 (n-4)\gamma\big) \xi_1\delta \xi_2\cr
&&\qquad +
\fft{(n-1)\omega\sigma_2}{32\pi} \big(4(n-1)\alpha + n\beta + 4\gamma\big)
\eta_1\delta \eta_2\,.
\eea
It follows from (\ref{wald2}) that the thermodynamical first law is
\bea
d\widetilde M &=& TdS - \fft{(n-2)\omega\sigma_1}{(n-1)32\pi}
\big(\kappa g^{-2} - 2(n-1)(\beta + n\alpha)+ 4 (n-4)\gamma\big) \xi_1 d \xi_2\cr
&&\qquad\quad -
\fft{(n-1)\omega\sigma_2}{32\pi} \big(4(n-1)\alpha + n\beta + 4\gamma\big)
\eta_1 d\eta_2\,.\label{adsfirstlaw}
\eea
The result is applicable for general asymptotic AdS black holes.  For asymptotically-flat black holes, a massive hair involve both the convergent as well as the divergent Yukawa falloffs.  The divergent falloff has to be excluded to maintain the asymptotic spacetime structure.  This requires a rather delicate fine tuning even if the no-hair theorem can be bypassed.  The resulting first law is then simply $dM=TdS$ since it requires both falloffs in the solution to form a conjugate pair to modify the first law.  The argument extends to the asymptotic AdS black holes as well, if one or both of the falloff terms are actually more divergent than $r^2$ and have to be excluded from the solution.

It is worth commenting that $\delta H_\infty$ has two parts in general, one is integrable associated with $\widetilde M$ and the other is non-integrable associated with $(\xi_1,\xi_2)$ and $(\eta_1,\eta_2)$.  The existence of the non-integrable part in $\delta H_\infty$ was interpreted as that the solution has no well-defined mass in \cite{Chow:2013gba}.  Even if this interpretation is valid, there is nevertheless a quantity which we label $\widetilde M$ here that has dimension of energy such that (\ref{adsfirstlaw}) is a correct mathematical equation for black holes at least at the classical level.  We shall not be precise or pedantic and continue to refer to the quantity $\widetilde M$ (or without the tilde) as ``mass" and call (\ref{adsfirstlaw}) as the ``first law of thermodynamics'' although whether it describes an actual physical thermal system is still questionable.

\subsection{Testing the first law}

Although we obtained the first law (\ref{adsfirstlaw}) from the linear analysis, we expect that it holds at the full non-linear level, with the understanding the expressions for $(\widetilde M, \xi_2,\eta_2)$ are subject to the non-linear modifications.  The similar situation occurs for the scalar and massive vector hairs studied previously \cite{liulu,Lu:2014maa,Liu:2014tra}. This should not be surprising, since the first law is bilinear in nature, and hence it must work at the linear level of a solution.  The non-linear contribution may augment the definitions of the charges, it will not change the coefficient of each term in the first law (\ref{adsfirstlaw}).

In this subsection, we shall nevertheless test the conjectured first law with some selected examples.

\subsubsection{Schwarzschild-like black holes}

We refer to the Schwarzschild-like black holes as those involving only the condensation of gravitons.  The metric functions in the large-$r$ expansion take the form
\be
h\sim f = g^2 r^2 + k - \fft{m}{r^{n-3}} + \cdots\,.
\ee
For these black holes, it is easy to verify that
\be
\delta H_\infty = \delta M\,,
\ee
where
\be
M=\fft{(n-2)\omega}{16\pi}\Big(\kappa -2n(n-1)\alpha g^2 - 2(n-1)\beta g^2 +
4(n-4) \gamma g^2\Big)m\,.\label{schwlikemass}
\ee
Thus for these black holes, the first law of thermodynamics is simply
\be
dM = T dS\,.\label{firstlaw1}
\ee
The mass formula (\ref{schwlikemass}) was previously obtained in \cite{Deser:2002rt,Deser:2002jk}.

For generic coupling constants, there is no known example of exact Schwarzschild-like black holes. It can be tricky even to find numerical solutions since it requires delicate fine tuning to get rid of the other modes; a proper no-hair theorem has yet been established however.  For some special class of parameters, the solutions do exist. The best known example is the Schwarzschild-like solution in the Gauss-Bonnet extension \cite{Boulware:1985wk,Cai:2001dz}.  A simpler example is to set the coupling constant $\gamma$ associated with the Riemann-squared term vanish. Einstein metrics of appropriate effective cosmological constant are then solutions. This implies that the theory admits AdS Schwarzschild black holes with
\be
h=f = g^2 r^2 + k - \fft{m}{r^{n-3}}\,,
\ee
and
\be
\Lambda_0=-\ft12 (n-1)(n-2) g^2 + \ft12 \kappa^{-1} g^4 (n-1)^2 (n-4) (\beta +n\alpha)\,.
\ee
It is then straightforward to verify that
\be
\delta H_{\infty} = \fft{(n-2)\omega}{16\pi} \Big(\kappa -2(n-1)g^2(\beta + n\alpha)\Big) \delta m\,.
\ee
This implies that we can define the mass of the black hole as
\be
M= \fft{(n-2)\omega}{16\pi} \Big(\kappa -2(n-1)(\beta + n\alpha)g^2\Big)m\,.
\ee
This reproduces the results in \cite{Deser:2002rt,Deser:2002jk}.  The entropy and the temperature are given by
\be
T=\fft{f'(r_0)}{4\pi}\,,\qquad S=\ft14 \omega r_0^{n-2} \Big(\kappa -2(n-1)(\beta + n\alpha) g^2\Big)\,.
\ee
Here $r=r_0$ is the event horizon with $f(r_0)=0$. The first law (\ref{firstlaw1}) is then straightforwardly satisfied, since both $M$ and $S$ are multiplied by a same constant factor from those in Einstein gravity. The thermodynamics of these types of black holes were previously studied in \cite{Nojiri:2001aj,Nojiri:2002qn}. Note that the mass and the entropy can vanish simultaneously, a characteristic phenomenon that occurs in critical gravities \cite{Lu:2011zk,Deser:2011xc}.

\subsubsection{With massive spin-2 hair}

We now consider a non-trivial example with non-vanishing massive spin-2 hair.  The massive scalar hair drops out naturally in the Einstein-Weyl combination (\ref{ew}).  We consider this example since it was already studied in \cite{Lu:2012xu} by numerical analysis that such a black hole indeed could exist.  However, its thermodynamical first law was not addressed in \cite{Lu:2012xu}. The solution is of Einstein-Weyl gravity in $n=4$ dimensions, with the parameter constrains (\ref{ew}),  $\gamma=-\ft23 \kappa g^{-2}$ and $\Lambda_0=-3g^2$.  Assuming $k=1$, the full large-$r$ expansions of $(h,f)$ at the asymptotic AdS region are given by
\bea
h &=& g^2 r^2 + \xi_1 r^{\fft32} +
\fft{5\xi_1^2}{12g^2}r + \fft{65 \xi_1^3}{864 g^4} r^{\fft12}  + 1+
\Big(\xi_2 + \ft{5}{24}(\fft{\xi_1}{g^2} + \fft{49 \xi_1^5}{10368g^8})\log r\Big)
\fft{1}{\sqrt{r}}\cr
&& -\Big(m - \ft{35}{432}( \fft{\xi_1^2}{g^4} + \fft{7\xi_1^6}{1152 g^{10}})\log r\Big)
\fft{1}{r} + \cdots\,,\cr
f &=& g^2 r^2 + \ft16 \xi_1 r^{\fft32} - \fft{5\xi_1^2}{432g^2} r -
\fft{5\xi_1^3}{576g^4} r^{\fft12} + 1 + \fft{5\xi_1^4}{1944 g^6} \cr
&& + \Big(-\fft{155\xi_1}{216g^2} + \fft{805 \xi_1^5}{746496 g^8} + \fft{5\xi_2}{6}
 + \fft{25 (10368 g^6\xi_1 + 49 \xi_1^5)}{1492992 g^8} \log r
\Big)\fft{1}{\sqrt{r}}\\
&&-\Big(
m - \fft{175 \xi_1^2}{324 g^4} + \fft{18235 \xi_1^6}{11943936 g^{10}} +
\fft{67 \xi_1 \xi_2}{72 g^2} + \fft{5 (404352 g^6 \xi_1^2 + 1519 \xi_1^6)}{
17915904 g^{10}}\log r \Big)\fft{1}{r} + \cdots\nn\,.
\eea
(The faster falloff dotted terms will not affect $\delta H_\infty$.) Substituting this into our formula (\ref{wald1}) and take the $r\rightarrow \infty$ limit, we find a convergent result
\be
\delta H_\infty = \delta \widetilde M - \fft{5\omega}{72\pi g^2} \xi_1 \delta \xi_2\,,
\ee
where
\be
\widetilde M= \fft{5\omega}{24\pi}\Big(- m + \fft{79 \xi_1^2}{1296 g^4}
- \fft{2 \xi_1 \xi_2}{9 g^2}
- \fft{27373 \xi_1^6}{26873856 g^{10}}
\Big)\,.
\ee
In four dimensions, $\omega=4\pi$ and hence the first law is
\be
d\widetilde M=T dS + \fft{5}{18g^2} \xi_1 d \xi_2\,.
\ee
This indeed reproduces the general first law (\ref{adsfirstlaw}) with the special parameters for this solution.

\subsubsection{With both massive scalar and spin-2 hair}

We now consider a black hole that involves all the hairs that extended gravity allows.
The parameters of the theory is given by
\be
n=5\,,\qquad \alpha=\fft{169}{744g^2}\kappa\,,\qquad
\beta = -\fft{64\kappa}{93 g^2}\,,\qquad \gamma=0\,,\qquad
\Lambda_0=-\fft{75g^2}{31}\,,\label{specpara}
\ee
which implies that
\be
\sigma_1=1\,,\qquad \sigma_2=2\,,
\ee
in the linearized analysis of section 4.1. Without loss of generality, we let $\kappa=1$, and find
\bea
h &=& g^2 r^2 + \xi_1 r^{\fft12} + 1 + \Big(\xi_2 + \fft{15\xi_1\eta_1}{16g^2} \log r\Big) \fft{1}{\sqrt{r}} -\fft{1}{2g^2}(\eta_1-\ft58 \xi_1^2)\fft{1}{r}\cr
&&+ \Big( \fft{5 \xi_1}{8 g^2} - \fft{35 \eta_1 \xi_2}{32 g^2}
 -\fft{28809 \eta_1^2 \xi_1}{7168 g^4} +
 - \fft{525 \eta_1^2 \xi_1}{512g^4} \log r\Big)\fft{1}{\sqrt{r^3}}\cr
&&-\Big(m + \fft{\eta_1 (4 \eta_1 - 75 \xi_1^2)}{160 g^4}\log r\Big) \fft{1}{r^2} + \cdots\,,\cr
f &=& g^2 r^2 + \eta_1 r + \ft38 \xi_1 r^{\fft12} + 1 + \fft{6\eta_1^2}{g^2} +
\Big(\ft58 \xi_2 + \fft{\eta_1\xi_1}{64g^2} + \fft{75 \eta_1 \xi_1}{128g^2}
\log r\Big) \fft{1}{\sqrt{r}}\cr
&& +\Big(\eta_2 + \fft{4032 \eta_1^3 - 128 \eta_1 g^2 - 75 g^2 \xi_1^2}{128 g^4} \log r\Big) \fft{1}{r} \cr
&&+\Big(\fft{5 \xi_1}{64 g^2} - \fft{117 \eta_1 \xi_2}{256 g^2} -\fft{22485 \eta_1^2 \xi_1}{8192 g^4} - \fft{1755 \eta_1^2 \xi_1}{4096 g^4} \log r\Big)
\fft{1}{\sqrt{r^3}}\cr
&&-\Big(m + \fft{176 \eta_1 \eta_2 - 27 \xi_1 \xi_2}{48 g^2} -
\fft{\eta_1 (1067456 \eta_1 + 377355 \xi_1^2)}{92160 g^4} +
\fft{757 \eta_1^4}{4 g^6}\cr
&&\qquad + \fft{\eta_1 (443520 \eta_1^3 - 13984 \eta_1 g^2 - 12075 g^2 \xi_1^2)} {3840 g^6} \log r \Big)\fft{1}{r^2} + \cdots\,.
\eea
Here again we consider the spherically-symmetric case corresponding to $k=1$.
Substituting this complicated large-$r$ solution into (\ref{wald1}), we find a simple outcome, namely
\be
\delta H_\infty = \delta \widetilde M - \fft{15\omega}{248\pi g^2}\xi_1 \delta \xi_2 + \fft{3\omega}{62\pi g^2} \eta_1 \delta \tilde \eta_2\,.
\ee
where $\widetilde M$ and $\tilde\eta_2$ are given by
\bea
M &=& \fft{15\omega}{31\pi}\Big(-m -\fft{3 (8 \eta_1 \tilde \eta_2 + 45 \xi_1 \xi_2)}{320 g^2} + \fft{\eta_1 (131648 \eta_1 - 10125 \xi_1^2)}{76800 g^4} +
\fft{63\eta_1^4}{320g^6}\Big)\,,\cr
\tilde\eta_2 &=& \eta_2 - \fft{135}{512g^2}\xi_1^2\,.
\eea
Thus, we see that not only the mass, but also the spin-0 hair is modified by the non-linear effects.   The resulting first law takes the form
\be
d\widetilde M = T dS + \fft{15\omega}{248\pi g^2}\xi_1 d \xi_2 -
\fft{3\omega}{62\pi g^2} \eta_1 d \tilde \eta_2\,.
\ee
This result is identical to the conjectured first law from the linearized analysis, after specializing to the parameters (\ref{specpara}).

\section{Lifshitz Black holes}

\subsection{Lifshitz vacua}

The AdS vacua in planar coordinates admits in its boundary Minkowski spacetime with one dimension less.  This geometry can be generalized to the Lifshitz spacetime \cite{Kachru:2008yh}
\be
ds^2=\ell^2 \Big(-r^{2z} dt^2 + \fft{dr^2}{r^2} + r^2 dx^i dx^i\Big)\,.
\ee
which is homogeneous but not Einstein.  The metric captures the different powers of dynamical scaling in the temporal and spatial directions, namely
\be
t\rightarrow \lambda^z t\,,\qquad x_i \rightarrow \lambda x_i\,,
\ee
exhibited in various condensed matter systems near fixed points.  The metric becomes AdS when $z=1$.  It turns out that Lifshitz vacua are rather common places in quadratically-extended gravities \cite{AyonBeato:2010tm}.  The solutions are simply constrained by the following two equations
\bea
\ell^2 &=& \fft{2}{\kappa} \Big( (2z^2 + 2(n-2)z + (n-2)(n-1))\alpha +
(z^2+n-2)\beta +2(z^2 -(n-2) z + 1)\gamma\Big)\,,\cr
\Lambda_0 &=& -\fft{1}{2\kappa \ell^4} \Big( \big(2z^2 + 2 (n-2) z + (n-2)(n-1)\big)^2\alpha \cr
&&\qquad\quad + (z^2 + n-2) \big(2z^2 + 2 (n-2) z + (n-2)(n-1)\big)\beta\cr
&&\qquad\quad + 2 \big( 2 z^4 - 2 (n-4) (n-2) z^2- 2 (n-2) (n-3)^2 z
+(n-1)(n-2)   \big)\gamma\Big)\,.\label{lifscons}
\eea

In this section, we shall study the thermodynamics of the Lifshitz black holes in extended gravities.  Many exact solutions were constructed in \cite{AyonBeato:2010tm,Lu:2012xu}. The thermodynamics of black holes constructed in \cite{Lu:2012xu} was also analysed there; the first law for a few limited examples of Lifshitz black holes \cite{AyonBeato:2010tm} were analysed using the quasi-local charge technique in \cite{Gim:2014nba}. Our general formula (\ref{wald1}) can be used to derive the first law in all cases.

The most general Ansatz that respects the isometry is given by
\be
ds^2 = - r^{2z} \tilde h(r) dt^2 + \fft{dr^2}{r^2 \tilde f(r)} + r^2 dx^i dx^i\,.
\ee
We shall from now on set without loss of generality $\ell=1$ for Lifshitz solutions. In other words,
\be
h=r^{2z} \tilde h\,,\qquad f=r^2 \tilde f\,.
\ee
We shall refer to $(\tilde h, \tilde f)$ as the thermalization factors of the Lifshitz vacua.  It turns out that in all examples of the exact solutions we consider in this paper, they are equal.

\subsection{A $z=6$ Lifshitz black hole in $n=4$ dimensions}

In four dimensions, the Gauss-Bonnet term is a total derivative and hence we can set $\gamma=0$ without loss of generality.  It was shown in \cite{AyonBeato:2010tm} that for the following choice of the coupling constants,
\be
\alpha=-\ft9{64}\kappa\,,\qquad \beta = \ft{25}{64}\kappa\,,\qquad
\Lambda_0=-\ft{51}{2}\,,\label{z6para}
\ee
the corresponding Lagrangian admits the following exact solution of $z=6$:
\be
ds^2 = -r^{12} (1-\fft{\mu}{r^4}) \,dt^2 + \fft{dr^2}{r^2(1-\fft{\mu}{r^4})} + r^2 dx^i dx^i\,.
\ee
In other words, we have $\tilde h=\tilde f = 1 -\mu/r^4$.  The solution describes a black hole that is asymptotic to $z=6$ Lifshitz vacua. The horizon of the black hole is located at $r=r_0\equiv \mu^{1/4}$.  Using the formulae in (\ref{generalts}), it is straightforward to calculate the temperature and entropy
\be
T=\fft{r_0^6}{\pi}\,,\qquad S=-\fft{15}{16}\omega r_0^2 \kappa\,.
\ee
Substituting the solution into (\ref{wald1}), we find
\be
\delta H_\infty = - \fft{15\omega \kappa}{32\pi} \mu\delta \mu
\ee
This suggests that the first law is simply
\be
dM=T dS\,,\label{z6fo}
\ee
where the mass is given by
\be
M=-\fft{15\omega\kappa}{64\pi}\mu^2\,.\label{z6mass}
\ee
That the parameter $\mu$ contributes quadratically to the mass formula is rather unusual, indicating that there is no massless graviton condensation in the solution.  It is thus instructive to study the general solution that involves also the massless graviton mode.

In order to study the issue thoroughly, we obtain first the linearized modes in the Lifshitz vacua.  Let
\be
\tilde h = 1 + h_1\,,\qquad \tilde f = 1 + f_1\,.
\ee
we find that at the linearized level, the solutions for $(h_1,f_1)$ are given by
\bea
h_1 &=& \xi_1 r^{\sqrt{66}-4} + \fft{\eta_1\log r + \eta_2}{r^4} -
\fft{m}{r^8} + \xi_2 r^{-\sqrt{66}-4}\,,\cr
f_1 &=& -\ft{1}{169}(53\sqrt{66} - 337)\xi_1 r^{\sqrt{66}-4} +
(\eta_2 - \ft1{5} \eta_1 + \eta_1\log r)\fft{1}{r^4}\cr
&&-\fft{24m}{19 r^8} + \ft{1}{169}(53\sqrt{66}+337)\xi_2 r^{-\sqrt{66}-4}\,.
\eea
As it will be demonstrated in section 5.4, the integration constants $(\xi_1,\xi_2)$ and $(\eta_1,\eta_2)$ correspond to the massive spin-2 and spin-0 modes respectively. It is clear that the $\xi_1$ term is divergent and has to be excluded from the black  hole solution.  To obtain the precise formula of the first law for the general solution, we perform the large-$r$ expansions and find
\bea
\tilde h &=& 1 + \fft{\eta_1 \log r + \eta_2}{r^4} +
\Big(-m + (\ft{114}{125} \eta_1\eta_2 + \ft{1244}{3125}\eta_1^2)\log r +
\ft{57}{125} \eta_1^2 (\log r)^2\Big)\fft{1}{r^8} + \cdots\,,\cr
\tilde f &=& 1 + \Big(\eta_2 - \ft15\eta_1 + \eta_1\log r\Big) \fft{1}{r^4} +
\Big(-\ft{24}{19} m - \ft{112}{2375} \eta_1\eta_2 + \ft{10343}{950000} \eta_1^2\cr
&&\qquad\qquad+(\ft{144}{125}\eta_1\eta_2 + \ft{1424}{3125}\eta_1^2)\log r + \ft{72}{125}
\eta_1^2 (\log r)^2 \Big)\fft{1}{r^8} + \cdots\,.
\eea
Substituting this non-linear solution into (\ref{wald1}), we obtain
\be
\delta H_{\infty}= \delta M - \fft{3\kappa}{128\pi} \eta_1\delta \eta_2\,,
\ee
where
\be
M=-\fft{625\omega\kappa}{1216\pi}\Big(m + \ft{57}{125} \eta_2^2 +
\ft{145809}{1250000}\eta_1^2 + \ft{4691}{12500}\eta_1\eta_2\Big)\,.\label{nonlinearmass}
\ee
The thermodynamical first law for the general $z=6$ Lifshitz black hole is thus
\be
dM = TdS + \fft{3\kappa}{128\pi} \eta_1 d\eta_2\,.
\ee
We see that the mass can have both linear $m$ and non-linear contributions. For the specific exact solution discussed earlier, we have $m=0=\eta_1$ and $\eta_2=\mu$.  The mass (\ref{nonlinearmass}) is indeed reduced to (\ref{z6mass}) and the first law is simply (\ref{z6fo}).

\subsection{A class of Lifshitz black holes in $n\ge 5$ dimensions}

In \cite{AyonBeato:2010tm}, a class of Lifshitz black holes were found in $n\ge 5$ dimensions, with the thermal factors
\be
\tilde h=\tilde f= 1 - \fft{\mu}{r^{\fft12(n+z-2)}}\,.
\ee
The solutions emerge for the following coupling constants
\bea
\alpha &=& \fft{\kappa}{2(n-3)(n-4)(n+z-2)^2P_4(z)}
\Big((n-2)^3 (n+2) (n^2-4n+36)
\cr
&& + 2(n-2) (3n^4-30n^3+124n^2-536n+1024) z \cr
&&+ (n-2) (19 n^3 - 330 n^2 + 2052 n -3640) z^2 +
 12 (11n^3-84n^2+196n -120) z^3 \cr
 &&- 3(19 n^2-168n+356) z^4 + 18 (3 n-4) z^5
 - 27 z^6\Big)\,,\cr
\beta &=& \fft{2(n^2-4+3z^2)\kappa}{(n-3)(n-4)(n+z-2)^2P_4(z)}\Big(
-(n-2)(n^3+2n^2-12n+24)\cr
&& + 2(n^3-4n^2+32n-80) z - 8(n^2-10) z^2 -
6(3n-4)z^3 + 9z^4\Big)\,,\cr
\gamma &=& \fft{(n^2-4+3z^2)\kappa}{2(n-3)(n-4)(n+z-2)P_4(z)}\Big(
(n-2)(n^2-4n+36) + (n-2)(5n-62)z\cr
&& + 3(9n-14) z^2 - 9z^3\Big)\,,\cr
\Lambda_0 &=& - \fft{n-2}{4P_4(z)}\Big((n-1)(n-2)^3(n+2) +
4(n-2)(n^3 - 2n^2 + 15n-62) z\cr
&&+2(n-2)(5n^2-73n+356) z^2 + 4 (19n^2-200n + 325) z^3 +
(197n-389)z^4\Big)\,,
\eea
where $P_4(z)$ is a fourth order polynomial in $z$, given by
\bea
P_4(z)&=&(n+2)(n-2)^3 + 4 (n-2) (n^2-n+30)z + 2(n-2) (5n-116)z^2\cr
&&+36(3n-5) z^3 - 27 z^4\,.
\eea
It is easy to see that the parameter $\mu$ is not associated with the condensation of the massless graviton which has the falloff $1/r^{n+z-2}$ \cite{Liu:2014dva}.  Substituting the solution into (\ref{wald1}), we find
\be
\delta H_\infty = \fft{(n-2)(n+2-3z)(n-2+3z)(n^2-4+3z^2)\kappa}{32\pi P_4(z)} \delta (\mu^2)\,.
\ee
The first law is then given by (\ref{z6fo}) with
\be
M=\fft{(n-2)(n+2-3z)(n-2+3z)(n^2-4+3z^2)\kappa}{32\pi P_4(z)} \mu^2\,.
\ee
The first law can be easily verified since the temperature and entropy, from (\ref{generalts}), are given by
\be
T=\fft{(n+z-2)}{8\pi} r_0^z\,,\qquad
S=\fft{(n+2-3z)(n-2+3z)(n^2-4+3z^2)\omega r_0^{n-2}}{4P_4(z)}\kappa\,,
\ee
with $\mu= r_0^{\fft12 (n+z-2)}$.

\subsection{General formula via linearized analysis}

We now consider the thermodynamical first law for a general Lifshitz black hole in quadratically extended gravity, as we did for the general AdS black hole in section 4. We first analyse $n=4$ dimensions, where we can, without loss of generality, set $\gamma=0$.  As we have mentioned earlier, we also set $\ell=1$, the Lifshitz solutions with general $z$ arise with the condition (\ref{lifscons}), which allows us to solve for $(\Lambda_0, \alpha)$ in terms of $\kappa$ and $\beta$. In terms of linearization, we find
\bea
h_1 &=& -\fft{m}{r^{z+2}} + \fft{\xi_1}{r^{\fft12(z+2-\sigma_1)}} + \fft{\xi_2}{r^{\fft12(z+2+\sigma_1)}} + \fft{\eta_1}{r^{\fft12(z+2-\sigma_2)}} + \fft{\eta_2}{r^{\fft12(z+2+\sigma_2)}}\,,\cr
f_1 &=& -\fft{z(z+2) m}{(z^2+2) r^{z+2}} + \fft{b_-\xi_1}{r^{\fft12(z+2-\sigma_1)}} + \fft{b_+\xi_2}{r^{\fft12(z+2+\sigma_1)}} + \fft{c_-\eta_1}{r^{\fft12(z+2-\sigma_2)}} + \fft{c_+\eta_2}{r^{\fft12(z+2+\sigma_2)}}\,,\cr
\sigma_1^2 &=& 3(3z^2-4z+4)\,,\qquad\sigma_2^2 =\fft{(11z^2 + 28 z+36) \kappa - 2z(z+2)^2(z-4)\beta}{3\kappa - 2z(z-4)\beta}\,,\cr
b_\pm &=& \fft{3z^3-z^2+11z-4 \pm(z^2+3z-1)\sigma_1}{2(2z+1)^2}\,,\qquad
c_\pm = \fft{z+4\pm\sigma_2}{2(z-1)}\,,
\label{lif4linear}
\eea
It can be demonstrated that in the Einstein-Weyl combination ($\beta + 3\alpha=0$), the modes associated with $(\eta_1,\eta_2)$ decouple, suggesting that they are the scalar modes. Setting $z=6$ then allows us to identify the massive spin-2 and scalar hairs in the solution discussed in the previous subsection. Furthermore, taking the parameters (\ref{z6para}) implies that $\sigma_2=0$, giving rising to the log modes in the previous subsection. As in case of the AdS black holes, we now substitute the linear solution into (\ref{wald1}) and obtain $\delta H_\infty$.  The result takes the same form as in (\ref{adswald1}), with $c_0$ now given by
\be
c_0=-\fft{(z+2)(z-1)^2\omega}{4\pi (z^2+2)}\beta\,.
\ee
The parameters $c_i$, with $i=1,2,3,4$, are complicated and not worth presenting.  Making an analogous Legendre transformation (\ref{legendre}), we find that the first law takes the simple form:
\be
d\widetilde M = TdS + \fft{z(z-4)\omega\,\sigma_2\,\beta}{8\pi(z-1)^2} \eta_1d\eta_2 -\fft{z(z-1)^2(z-4)\omega\,\sigma_1\,\beta}{8\pi(2z+1)^2} \xi_1 d\xi_2\,.
\ee
For the same argument we gave in the case of AdS black holes, we expect that the first law is valid even though it was derived from the linear analysis.  It is understood however that the mass and the charges can acquire some non-linear modifications in a specific solution.  It is of interest to note that the parameter $\kappa$ enters the first law via $\sigma_i$, and there is a smooth $\kappa\rightarrow 0$ limit .   There is no smooth limit of $\beta\rightarrow 0$, and in fact there is no Lifshitz solution when $\beta=0=\alpha$.  Also note that for non-vanishing $\eta_i$, there is no smooth $z=1$ limit.

The situation becomes much more complicated for $n\ge 5$ dimensions.  Although the linear solutions take the similar form as (\ref{lif4linear}), $\sigma_1^2$ and $\sigma_2^2$ now satisfy a non-factorable quadratic expression (quartic in $\sigma_1$ or $\sigma_2$).  The expression for the general first law becomes rather horrid and we do not find it instructive to present here.

\subsection{Black holes with vanishing mass and entropy}

There are two more classes Lifshitz solutions with $n>4$ which were found in \cite{AyonBeato:2010tm}. One is given by
\be \tilde h (r)=\tilde{f}(r)=1-\frac{m}{r^{2z-2}} \ee
and the coupling constants are:
\bea \alpha&=&\kappa(8-10z+20z^2-6z^3+n^3z(2z-3)\cr
           &&- n^2(3z^3+2z^2-14z+1)+n(9z^3-11z^2-19z+2))/P(z)\,,\cr
     \beta&=&\kappa(n-1)(n^3z-n^2(8z^2+z-2)\cr
           &&+2n(6z^3+3z^2-z-2)+8(-3z^3+z^2+4z-2))/P(z)\,,\cr
     \gamma&=&\frac{\kappa(n-1)(n-2z+2)}{4z(n-4)(n-3)(n-z)}\,,\qquad
     \Lambda=-\frac{(z-1)(2+n^2+n(z-3)-z^2)}{2\ell^2(n-z)}\,,
\eea
where $P(z)=2z(n-4)(n-3)(n-2)(n-z)(z-1)(n-4+3z)$. The other is given by
\be
\tilde h(r)=\tilde{f}(r)=1-m r^z\,.
\ee
The corresponding coupling constants are
\bea \alpha&=&-\frac{\kappa(5+4z-3z^2-2n(z+1))}{2z(n-4)(n-3)(n+2z-2)}\,,\quad
     \beta=-\frac{\kappa (2+n^2-8z+6z^2+n(4z-3))}{z(n-4)(n-3)(n+2z-2)}\,,\cr
     \gamma&=&-\beta/4\,,\qquad
      \Lambda=\frac{z(6-5n+n^2-8z+4nz+2z^2)}{4\ell^2(n+2z-2)}\,.
\eea
This solution is of less interest since it requires $z<0$. The temperatures of the two solutions are respectively given by
\be
T=\frac{(z-1)r_0^z}{2\pi}\ (\mbox{for}\ z>1 )\,,
\qquad T=-\frac{z r_0^z}{4\pi}\ (\mbox{for}\ z<0)\,.
\ee
However, both of these two solutions have vanishing entropy and black hole mass ($\delta H_\infty=0$), and thus their first law is trivially satisfied.  This phenomenon is analogous to that occurs in black holes of critical gravities \cite{Lu:2011zk,Deser:2011xc}.

\section{Constructing new black holes}

Using the formula (\ref{wald1}), we have derived/rederived the thermodynamical first laws of Lifshitz black holes constructed in literature for quadratically-extended gravities.  In this section, we shall construct some new solutions and obtain their first laws.

\subsection{Exact Lifshitz black holes in quadratic gravity}

In this subsection, we consider purely quadratic gravity, corresponding to setting $\kappa=0$ in the general action (\ref{genaction}).  This has the effect of turning off the bared cosmological constant in the Lagrangian also.  We focus on the construction of exact Lifshitz black holes in this theory. It turns out that the existence of Lifshitz spacetime requires that $\gamma=0$. The Lifshitz exponent $z$ is determined quadratically in terms of the ratio $\beta/\alpha$:
\be
\fft{\beta}{\alpha}=-\fft{2z^2+2(n-2)z + (n-1)(n-2)}{z^2+n-2}\,.
\ee
It is worth pointing out that when $\gamma=0$, Einstein metrics including AdS Schwarzschild black holes, become solutions of the theory, which we have discussed in section 4.3.1.  Thus in this section, we shall avoid the repetition by requiring that $z\ne 1$. In four dimensions, when $\beta=-3\alpha$, the theory becomes conformal gravity, up to a Gauss-Bonnet term which is a total derivative.  In this case, one has $z=0$ and $z=4$.  The exact Lifshitz black holes were constructed in \cite{Lu:2012xu}.  It turns out that the $z=4$ solution can be generalized to higher dimensions to $z=n$, corresponding to setting
\be
\beta=-\fft{5n-2}{n+2}\alpha\,.
\ee
We find a class of exact solutions
\be
ds^2 = -r^{2n} \tilde f dt^2 + \fft{dr^2}{r^2\tilde f} + r^2 dx^i dx^i\,,\qquad
\tilde f = 1 - \fft{m}{r^{2n-2}}\,.
\ee
It is easy to see that $m$ is associated with the massless graviton mode and it contributes linearly to the mass.  Indeed, substituting the solutions into (\ref{wald1}), we find
\be
\delta H_\infty = \fft{(n-1)^2(n-2)^2 \omega \alpha}{2(n+2)\pi}
\delta m\,,
\ee
which implies that
\be
M=\fft{(n-1)^2(n-2)^2 \omega \alpha}{2(n+2)\pi}m\,.
\ee
The temperature and the entropy (\ref{generalts}) are given by
\be
T=\fft{n-1}{2\pi}\,r_0^n\,,\qquad S=\fft{2\alpha (n-2)(n-1)^2 \omega}{n+2}\, r_0^{n-2}\,,
\ee
with $m=r_0^{2n-2}$.  It is easy to verify that the first law $dM=TdS$ is indeed satisfied.

\subsection{Charged Lifshitz black holes}

In this subsection, we study extended gravities (\ref{genaction}) coupled to a Maxwell field with the Lagrangian $\sqrt{-g} L_A$, where
\be
L_A=-\ft14 F_{\mu\nu} F^{\mu\nu}\,.
\ee
We consider solutions carrying electric charges, with the Ansatz
\be
A=A_t(r) dt\,.
\ee
Its contribution to the variation of the Hamiltonian in the Wald formalism is given by \cite{Liu:2014tra,Liu:2014dva}
\be
\delta H^{(A)}=-\frac{\omega}{16\pi}r^{n-2}\sqrt{\frac{h}{f}}\, \Big(\frac{f}{h}A_t\delta A_t'+\frac 12 A_t A_t'(\frac{\delta f}{h}-\frac{f\delta h}{h^2})\Big)\,.\label{waldA}
\ee
For the gauge choice where $A_t$ vanishes on the horizon, it can be shown that $\delta H^{(A)}_+=0$, whilst $\delta H_\infty^{(A)}$ typically gives an non-integrable part, i.e.~$\sim\Phi dQ$.

\subsubsection{$n=4$ dimensions}

In \cite{Lu:2012xu}, Lifshitz black holes with $z=0$ and $z=4$ were constructed in four dimensional conformal gravity, corresponding to setting $\kappa=0$.  In this subsection, we obtain charged Lifshitz black holes with the same $z$ in conformal gravity coupled to the Maxwell field that is also conformal in four dimensions. We begin with the $z=0$ case. We find that the thermal factors and the gauge potential for the charged solution are given by:
\be
\tilde h (r)=\tilde{f}(r)=1+\frac{c}{r^2}+\frac{c^2-\tilde{q}^2}{3r^4},\qquad A=q(\frac{1}{r_0^2}-\frac{1}{r^2})dt\,,
\ee
where $\tilde{q}^2=-\frac{3q^2}{16\beta}$ and $r_0$ is the horizon parameter which is the largest root of the equation $\tilde{f}(r)=0$. Here we have chosen a gauge that the vector vanishes on the horizon. By simple calculation, we find
\be
r_0^2=\frac{-3c+\sqrt{3(4\tilde{q}^2-c^2)}}{6}\,.
\ee
To have horizon, we require $ -2\tilde{q}\leq c<\tilde{q}$. The temperature and entropy are given by:
\be
T=\frac{1}{\pi(1-\frac{\sqrt{3}c}{\sqrt{(4\tilde{q}^2-c^2)}})},\qquad S=\frac{\omega \beta}{3}(c+\sqrt{3(4\tilde{q}^2-c^2)})\,.
\ee
The solution becomes extremal when $c\rightarrow -2\tilde{q}$.
The total electric charge and potential can also be calculated using standard technique:
\be
Q=-\frac{1}{16\pi}\int {*F}=\frac{\omega q}{8\pi},\qquad \Phi=A_t|_{r\rightarrow \infty}=\frac{q}{r_0^2}\,.
\ee
From (\ref{wald1}), we obtain
\be
\delta H_\infty=\frac{\beta\omega}{3\pi}\delta c-\frac{\omega}{8\pi r_0^2}q\delta q\,.
\ee
Therefore, we should define the mass of the black hole and write the first law as:
\be
M=\frac{\beta\omega}{3\pi} c,\qquad dM=T dS+\Phi d Q\,.
\ee
It is worth mentioning that that the black hole mass here can also be calculated independently from the Noether charge 2-form  $M=\int Q_{(2)}$ given in \cite{Lu:2012xu}.
Thus every quantity in the first law can be computed independently and the first law can be verified immediately.

Now let us consider the $z=4$ case, for which, we find
\be
\tilde h(r)=\tilde{f}(r)=1+\frac{c}{r^2}+\frac{c^2-\tilde{q}^2}{3r^4}
+\frac{d}{r^6},\qquad A=q(r^2-r_0^2)dt\,,
\ee
where we have also $\tilde{q}^2=-\frac{3q^2}{16\beta}$ and $\tilde f(r_0)=0$. This is a three order equation of
$r^2$. It has one real root or three real roots, depending on the discriminant $\Delta$ is positive or negative:
\be
\Delta=(c^3-27d-3c\tilde{q}^2-2\tilde{q}^3)(c^3-27d-3c\tilde{q}^2+2\tilde{q}^3)
\ee
The temperature and entropy can be obtained as
\be
T=\frac{(c+3r_0^2)^2-\tilde{q}^2}{6\pi},\qquad S=-\frac{2\beta\omega}{3}(c+3r_0^2)\,.
\ee
The total electrical charge is given by:
\be
Q=\frac{\omega q}{8\pi}\,.
\ee
However, the electric potential cannot be defined in the usual way since the potential $A_t$ diverges in the asymptotical limit.  This phenomenon was also seen in \cite{Liu:2014dva}, and it would become hopeless to consider the first law in the conventional approach.

As in the case that was discussed in \cite{Liu:2014dva}, we can nevertheless substitute the solution into (\ref{wald1}), and a convergent result emerges:
\be
\delta H_\infty=-\frac{\omega\beta}{9\pi}[-9\delta d+(c^2-\tilde{q}^2)\delta c+6r_0^2\tilde{q}\delta \tilde{q}]\,.\label{z4deltaH}
\ee
The result implies that a better and more universal way of defining the electric potential, after fixing the gauge that $A_t(r=r_0)=0$, is simply to take the integration constant other than the charge parameter, namely
\be
\Phi=-q r_0^2\,.
\ee
One may take a view that since the gauge field $A$ diverges in the asymptotical limit, the electric potential needs to be defined properly:
\be
\Phi=A_t^{(reg)}|_{r\rightarrow \infty},\qquad A_t^{(reg)}=A_t-q r^2\,.
\ee
From (\ref{z4deltaH}), it is necessary to introduce a new pair thermodynamical variables which we call ``massive spin-2 hair'':
\be
\Xi=\frac{\omega\beta}{3\pi}c,\qquad \Psi=\frac{c^2-\tilde{q}^2}{3}\,.
\ee
Then the mass of the black hole and the first law can be written as:
\be
M=\frac{\omega\beta d}{\pi},\qquad dM=T dS+\Phi d Q+\Psi d\Xi\,.
\ee
It is worth remaking that the thermodynamics of the $z=4$ neutral solution was studied in \cite{Lu:2012xu}.  However, owing to the lack of proper definition of the mass and massive spin-2 hair, the first law was not properly derived.

\subsubsection{Higher dimensions $n\ge5$}

In $n\ge 5$ dimensions, we obtain a new class of exact black hole solutions with electric charges. The thermal factors and the gauge potential are given by
\be
\tilde h (r)=\tilde{f}(r)=1-\frac{\lambda q^2}{r^{2n-4}}\,,\qquad A=\sqrt{\kappa}\, q\Big(\frac{1}{r_0^{n-z-2}}-\frac{1}{r^{n-z-2}}\Big)dt\,,
\ee
where $\lambda$ is a pure constant given by:
\be \lambda=\frac{z((n-4)(n-2)^2+n(n-2)z-3(n-2)z^2+z^3)}{\kappa
(n-2)(2(n-2)^2(n-1)-(n-2)(7n-15)z+2z^3)}\,.
\ee
The coupling constants $(\alpha,\beta, \gamma)$ and the cosmological constant are all fixed, given by:
\bea
\alpha &=& -\frac{\kappa\, N(z)}{4z(n-2)(n-3)(n-4)(n-z-2)((n-4)(n-2)+2(n-2)z-z^2)}\,,\cr
N(z) &=& 2(n-4)(n-2)^3+(n-2)(-4-5n+3n^2)z-2(n-2)(3-8n+3n^2)z^2\cr
&&+ (20-25n+7n^2)z^3-2(n-2)z^4\,,\cr
\beta &=& \frac{\kappa\, K(z)}{4z(n-2)(n-3)(n-4)(n-z-2)((n-4)(n-2) +2(n-2)z-z^2)}\,,\cr
K(z) &=& \Big((n-1)(n-2)-4(n-2)z+2z^2\Big)\Big(2(n-1)(n-2)(n-4)\cr
&&+ (20-25n+7n^2)z-4(n-2)z^2\Big)\,,\cr
\gamma &=&-\frac{\kappa\Big((n-1)(n-2)-4(n-2)z+ 2z^2\Big)}{4z(n-3)(n-4)(n-z-2)}\,,\cr
\Lambda_0 &=& -\frac{(n-2)\Big((n-1)(n-2)-(n-2)z-z^2\Big)}{2(n-z-2)}\,.
\eea
From (\ref{wald1}), we have
\be
\delta H_\infty=-\frac{\omega(n-z-2)\kappa\, q\delta q}{16\pi \ r_0^{n-z-2}}\,.
\ee
The Wald equation $\delta H_\infty=T \delta S$ then implies
\be
TdS+\Phi dQ=0\,.\label{masslessfo}
\ee
Here the electric potential and total electric charge are given by
\be
\Phi=\frac{\sqrt{\kappa}\, q}{r_0^{n-z-2}},\qquad Q=-\frac{\omega (n-z-2)\sqrt{\kappa}\,q}{16\pi}\,.
\ee
To verify that above first law is indeed satisfied by our solutions, we obtain the temperature and entropy from (\ref{generalts}), given by
\be
T=\frac{(n-2)r_0^z}{2\pi},\qquad S=-\frac{\omega (n-z-2)\kappa}{8(n-2)\lambda}\, r_0^{n-2}\,.
\ee
It is then immediately verifiable that the first law (\ref{masslessfo}) is indeed satisfied.  The unusual looking massless first law (\ref{masslessfo}) was also observed previously \cite{Zingg:2011cw,Liu:2014dva}.

\section{Conclusions}

In this paper, we considered Einstein gravities with a cosmological constant extended with quadratic curvature invariants in general dimensions.  The theories admit a variety of black holes that are asymptotically flat, AdS and Lifshitz spacetimes.  These black holes may involve condensations of massive scalar and spin-2 modes, in addition to the usual massless graviton modes.  Whilst black holes involving only the massless graviton mode has the usual first law of thermodynamics, the situation becomes very complicated when solutions involve additional hairs.

We focused our discussion on the static black holes with spherical/toric/hyperbolic isometries.  The most general metric Ansatz is given in (\ref{metricansatz}), involving two metric functions $(h,f)$. We applied the Wald formalism and derive an explicit formula (\ref{wald1}) in terms of $h$ and $f$ for the variation of the Hamilitonian. The formula can be readily used to derive the thermodynamical first law for any black holes within the class of (\ref{metricansatz}) in extended gravities.  We used the formula to derive or re-derive the first laws for wide classes of black holes of extended gravities in literature.  In addition, we constructed many new black holes and derived their first laws.

We would like to emphasize the point again that has already been addressed under (\ref{adsfirstlaw}). Namely, whether the quantities we labelled as $M$ or $\widetilde M$ can be called the proper mass of the black holes and/or whether the equation (\ref{adsfirstlaw}) can be actually promoted as the thermodynamical first law where $\xi_i$ and $\eta_i$ are true thermal variables in an actual physical thermal setup are both debatable. What can be said definitely at this point is that $M$ and $\widetilde M$ are quantities that have dimension of energy and the equation like (\ref{adsfirstlaw}) for AdS black holes, and the similar one for Lifshitz black holes derived from (\ref{wald1}) are mathematically correct and non-trivial equations at the classical level for all static black holes in quadratically-extended gravities considered in this paper.  No matter how they should be interpreted eventually, they surely capture an important property of these black holes.

Our formula (\ref{wald1}) provides a powerful and useful tool to derive the first law of static black holes in quadratically extended gravities .  It is of great interest to generalize the results to include rotations.

\section*{Acknowledgement}

Z.-Y.~Fan is supported in part by NSFC Grants NO.10975016, NO.11235003 and NCET-12-0054; The work of H.L.~is supported in part by NSFC grants NO.11175269, NO.11475024 and NO.11235003.

\end{document}